\documentclass{llncs}
\usepackage{graphicx}

\usepackage{t1enc}
\usepackage[cp1250]{inputenc}

\usepackage{latexsym}
\usepackage{amssymb}
\usepackage{tabularx}
\usepackage{amsfonts}
\usepackage{listings} 

\usepackage{colortbl}
\usepackage{xcolor}
\usepackage{pst-all} 
\usepackage{pst-poly}
\newpsobject{showgrid}{psgrid}{subgriddiv=1,griddots=10,gridlabels=6pt}
\usepackage{fancybox}
\usepackage{wrapfig}

\usepackage{algorithm}
\usepackage{algpseudocode}

\newcommand{\con}{\wedge} 
\newcommand{\dis}{\vee} 
\newcommand{\alw}{\Box} 
\newcommand{\imp}{\Rightarrow} 
\newcommand{\som}{\Diamond} 

\newcommand{\theo}{\mbox{$|\!-\,$}}

\title{Generating Logical Specifications from Requirements Models for Deduction-based Formal Verification}
\author{Rados{\l}aw Klimek}
\date{January 2014}
\institute{AGH University of Science and Technology,\\
           al.\ A.\ Mickiewicza 30, 30-059 Krakow, Poland\\
           \email{rklimek@agh.edu.pl}
           }

\begin{document}
\maketitle

\begin{abstract}
The work concerns automatic generation of logical specifications from requirements models.
Logical specifications obtained in such a way can be subjected to formal verification using deductive reasoning.
Formal verification concerns correctness of a model behaviour.
Reliability of the requirements engineering is essential for all phases of software development processes.
Deductive reasoning is an important alternative among other formal methods.
However, logical specifications, considered as sets of temporal logic formulas,
are difficult to specify manually by inexperienced users and this fact can be regarded as
a significant obstacle to practical use of deduction-based verification tools.
A method of building requirements models using some UML diagrams,
including their logical specifications, is presented step by step.
Organizing activity diagrams into predefined workflow patterns
enables automated extraction of logical specifications.
The crucial aspect of the presented approach is integrating the requirements engineering phase and
the automatic generation of logical specifications.
A system of the deduction-based verification is proposed.
The reasoning process could be based on the semantic tableaux method.
A simple yet illustrative example of the requirements elicitation and verification is provided.

\end{abstract}

\section{Introduction}
\label{sec:introduction}


Software modeling enables better understanding of the domain problem and the system under development.
Requirements engineering is an significant part of software modeling.
Requirements elicitation should lead into a coherent structure of
requirements and have fundamental impact on software quality and costs.
Thinking of requirements must precede the analysis, design, and code generation acts.
Requirements models are descriptions of delivered services
in the context of operational constraints.
Identifying software requirements of the system-as-is,
gathering requirements and formulation of requirements by users
allows defects to be identified earlier in a software life cycle.

UML, i.e.\ the Unified Modeling Language~\cite{Booch-Rumbaugh-Jacobson-1999,Pender-2003},
which is ubiquitous in the software industry can be a powerful tool for
the requirements engineering process.
Use cases are central to UML
since they strongly affect other aspects of the modeled system
and, after joining the activity diagrams,
may constitute a good vehicle to discover and write down requirements.
Temporal logic is a well established formalism which allows to describe properties of reactive systems,
also visualized in UML.
The semantic tableaux method,
which is a proof formalization for assessing logical satisfiability,
and which might be descriptively called ``satisfiability trees'',
seems intuitive and may be regarded as goal-based formal reasoning.

Formal methods enable precise formulation of important artifacts
arising during software development and help eliminate ambiguity.
There are many examples of successful application for formal methods,
e.g.~\cite{Abrial-2007},
and there are challenges for the future~\cite{Hoare-Misra-2005}.
There are two well established approaches to formal reasoning and
system verification~\cite{Clarke-Wing-etal-1996}.
The first is based on state exploration (``model checking'')
and the second is based on deductive reasoning.
However, model checking is an operational rather than analytic approach~\cite{Clarke-etal-1999}.
Deductive inference enables the analysis of infinite computation sequences.
On the other hand,
one important problem of the deductive approach is the lack of automatic methods for obtaining
logical specifications considered as sets of temporal logic formulas.
Even in the case of an average-size system,
it consists of many formulas and
it is not possible to build a logical specification manually,
which can be recognized as a major obstacle to untrained users.
Thus, the automation of this process seems particularly important.
Moreover, application of the formal approach to the entire requirements engineering phase may increase
the maturity of requirements models.

The main issue addressed in the work is a theoretical idea that is verified by
finding a workable and comprehensive solution and analyzing a simple yet illustrative and real example.
Theoretical problem relates to
the generation of logical specifications for requirements models and
the application of deductive approach to formal verification of a model behaviour.
Logical specifications are always important for formal methods since
they constitute proper, formal and representative images/views of the developed model/system.
Consequently,
they can be used in many phases of the software development cycle,
as formal methods offer numerous application possibilities~\cite{Woodcock-etal-2009}.
On the other hand, deductive reasoning plays an important role in the formal approach
as a ``top-down'' and sustainable way of thinking,
with reasoning moving from a more general facts to the more specific ones
to reach a logically certain conclusion.
Let us consider some arguments in favor of a deductive approach.
\begin{itemize}
  \item The first and the strong argument is the fact that deductive reasoning enables
        analyzing infinite sequences of computations.
  \item Another argument is naturalness and common use of deductive reasoning in everyday life.
        It also dominates in scientific works where obvious deductions are ubiquitous and
        represent a rational thought sequence that moves linearly from the premises to
        the conclusion what resembles our normal reasoning.
  \item A kind of informal argument is an analogy between natural languages and logical approach,
        that is the strict application of formal grammatical rules,
        although not necessary, however raises the quality of statements in
        a natural language, like, by analogy, there is no doubt that
        the strict application of logical rules for reasoning increases the quality of verification procedures and makes them more reliable.
\end{itemize}
Logical specifications and deductive reasoning might help to truly understand software models,
as they should be understood, what is fundamental in obtaining trustworthy designs
as to understand that realizing a good model is more important than
producing code~\cite{Hinchey-etal-2008}.
Hence, logic enables evaluation of arguments,
i.e.\ it contains methods and procedures for checking the ``reliability'' of arguments.
Arguments are statements consisting of evidence and a conclusion.
Evidence statements are premises while the conclusion must follows from these premises.
Thus, the work is also focused on the deduction-based formal verification,
i.e.\ arguing from the general to the particular.
It seems that workflow-oriented requirements models are suitable for this kind of verification.
Informally speaking,
workflows are focused on processes and are not disturbed by data flows.

\subsection{Motivations and contributions}
The motivation for this work is the lack of tools for
automatic extraction of logical specifications from software models.
Another motivation, which is associated with the previous one,
is the lack of tools and practical applications of deductive methods for
formal verification of requirements models.
However, requirements models built using use cases and
activity diagrams seem to be suitable for such an extraction process.
All of the above mentioned aspects of the formal approach
seem to be an intellectual challenge in software engineering.

The contribution of the work is a method for building formal requirements models,
including their logical specification, based on some UML diagrams.
A complete deduction-based system which enables the automated and
formal verification of requirements models is proposed.
The correctness of a model behaviour is considered.
Another contribution is a method for
automating the generation of logical specifications.
The generation algorithm for some workflow patterns is presented.
Although the work is not based on any particular method of reasoning,
i.e.\ generated logical specifications can be used for many purposes
and reasoning engines,
the semantic tableaux method for temporal logic is suggested.
The proposed generating method is characterized by the following advantages:
introducing workflow patterns as logical primitives to requirements engineering and logical modeling,
scaling up to real-world problems,
i.e.\ migration from small models to real problems in the sense that they are having
more and more predefined patterns and more nesting expressions,
and logical patterns once defined,
e.g.\ by a logician or a person with good skills in logic,
and widely used,
e.g.\ by analysts and engineers with less skills in logic.
All these factors are discussed in the work and summarized in the last Section.

\subsection{Related works}
There is a large volume of published works describing the role and importance of requirements engineering,
as well as some conferences and journals are dedicated to this subject.
There is an unambiguous relationship between
the quality of the requirements engineering phase and the quality of developed system.
Some fundamental works on requirements engineering are published,
c.f.\ works by Sommervile~\cite{Sommerville-Kotonya-1998}, van Lamsweerde~\cite{vanLamsweerde-2009},
or by Pohl~\cite{Pohl-2010}
which are comprehensive studies of many fundamentals of this area.
Work by Chakraborty et al.~\cite{Chakraborty-etal-2010} discusses
some social processes associated with requirements engineering.
Work edited by Yu et al.~\cite{Yu-etal-2011} discuses some aspect of
social modeling for requirements engineering including
modeling framework,
applications in security/privacy,
incorporating and evaluating social modeling, etc.
In work by Winkler and Pilgrim~\cite{Winkler-Pilgrim-2010} the problem of
traceability in the requirements engineering context is discussed.
Traceability is understood as the ability to follow the life of software artifacts.
Work is a review of research and practice
identifying commonalities and differences in these areas.
Work by Cao and Ramesh~\cite{Cao-Ramesh-2008} discusses requirements engineering
in agile development, and some real cases are considered.
This approach is suitable for rapidly changing business environment
and differs from traditional approach, thus,
it should be used when developing unambiguous
and complete requirement specifications is impossible.

There are many approaches for building and analysis requirements models.
In the work by Rauf et al.~\cite{Rauf-etal-2011}, a method for extracting
logical structures from text documents is presented,
however, this solid work concerns rather recognition instances of use cases, business rules, etc.,
and discovered relations are not understood in a formal/logical way, i.e.\ in terms of logical specifications.
Work by Kazhamiakin et al.~\cite{Kazhamiakin-2004} discusses
a method based on formal verification of requirements, as well as a case study on web services is discussed,
however, linear time logic and model checking is used.
Blanc et al.~\cite{Blanc-etal-2008} propose a kind of formalism as sequences of elementary constructions
that originate from different models, and are considered uniformly in the work,
then the logic-based analysis of inconsistency using the Prolog engine is performed.
The work constitute an interesting logic-based approach and distinctive dissimilarity to other works/approaches.
Work by Nikora and Balcon~\cite{Nikora-Balcom-2009} provides a method for identifying and discovering
temporal properties contained in a natural language as requirements.
These requirements are specified as temporal logic patterns.
Some machine learning techniques are used.
Properties can be converted to finite state automata and analyzed using model checking techniques.
Work by Smith and Havelund~\cite{Smith-Havelund-2008} is another work providing tools,
that graphically cover formal requirements, and enabling verification using
the model checking techniques.

Many works concern formalization, and verification, of the UML diagrams that are used in the work.
Hurlbut~\cite{Hurlbut-1997} provides a~very detailed
survey of selected issues concerning formalization of use cases.
The informal character of scenario documentation
implies several difficulties in reasoning about the system
behavior and validating the consistency between the diagrams and
scenario descriptions.
Work by Barrett et al.~\cite{Barrett-etal-2009}
proposes formal definiton of syntax and semantics of use cases
to enable modeling of use cases,
detecting their inconsistencies and conflicts.
Some illustrative examples are presented.
Zhao and Duan~\cite{Zhao-Duan-2009} shows formal analysis of use cases;
however, the Petri Nets formalism is used.
The solid work by Eshuis and Wieringa~\cite{Eshuis-Wieringa-2004} addresses
the issues of activity diagram workflows but the goal is to
translate diagrams into a~format that allows model checking.
Then, propositional requirements are checked against the input model.
Some examples are provided.
In the work by Cabral and Sampaio~\cite{Cabral-Sampaio-2008},
a method for automatic generation of
use cases and a proposed subset of natural languages,
however, for the algebraic specification is introduced.
Work by Rossi et al.~\cite{Rossi-etal-2004}
provides interesting and comprehensive formalization of state machines using temporal logic,
however, only state diagrams are discussed which are not considered in the work.
There is a variety of formalisms used in these area,
however, they are not discussed widely in the work.

In the comprehensive and general work by Shankar~\cite{Shankar-2009} automated deduction for verification
using symbolic logical reasoning is widely discussed.
There are considered satisfiability procedures, automated proof search, and variety of application
in the case of propositional and fragments of first-order logic.
However, even though the work contains a survey of symbolic reasoning,
modal and temporal logics are not considered widely.
As it is already cited in work~\cite{Rossi-etal-2004},
the statement by Chomicki and Saake taken from~\cite{Chomicki-Saake-1998},
``Logic has simple, unambiguous syntax and semantics.
It is thus ideally suited to the task of specifying information systems''.
Logic offers also many possibilities of applications,
i.e.\ specification, verification, synthesis, and programming~\cite{Galton-1992}.
``Logic is the glue that binds together methods of reasoning, in all domains'',
thus
``we need a style of logic that can be used as a tool in every-day work''~\cite{Gries-Schneider-1993}.
Moreover,
``automated deduction tools can be used in a variety of ways in formal verification in
applications ranging from modeling requirements and capturing program semantics to
generating test cases''~\cite{Shankar-2009}.
In work~\cite{Clarke-Wing-etal-1996} some examples of existing theorem provers,
i.e.\ when both the system and desired properties are expressed as logical formulas,
with a different degree of automation,
are briefly discussed.
In the end of Section~\ref{sec:logical-background},
some works discussing experimental results with automated theorem provers using temporal logic are presented.

Summing up, there are many works in the domain of requirements engineering,
and works in the area of the formal approach for the UML-based requirements engineering
but there is a lack of works for extracting logical specifications from requirements models and
deduction-based formal verification with temporal logic as well as the semantic tableaux method for UML-based requirements models.
This work is an extended version of the work~\cite{Klimek-2013-sefm}.
The major differences are that some new formal notions are introduced,
the generation algorithm is improved and expressed in a more formal way,
as well as more widely discussed,
the main example of the work is extended both in the case of modeling and reasoning.

\subsection{Organization}

The rest of the work is organized as follows.
The procedure and guidelines for the construction of formal requirements models are
introduced in Section~\ref{sec:methodology}.
The presented procedure contains both manually and automatically performed steps.
The problem of identifying atomic activities using use case diagrams and scenario is discussed
in Section~\ref{sec:use-cases-scenarios}.
The appropriate example is shown.
Logical background which are temporal logic and logical inference using the semantic tableaux method
are discussed in Section~\ref{sec:logical-background}.
Temporal logic is an established standard for the specification and verification of reactive systems
and the semantic tableaux method is a natural and valuable method of inference.
The deduction system and its architecture is also proposed.
The system enables formal verification of requirements models.
Workflow patterns for activity diagrams are introduced in Section~\ref{sec:modeling-activities}.
They are treated as (logical) primitives which allow to automate
the entire process of generating logical specifications.
Launched real example is continued.
The algorithm for generation logical specifications is proposed in Section~\ref{sec:generating-logical-specifications}.
Introduced workflow are predefined in terms of temporal logic formulas.
Some properties of the proposed generation algorithm is discussed in the end of the Section.
The general example for the requirements model considered in previous Sections is continued in Section~\ref{sec:reasoning-verification}
by the generating logical specifications for the model and formal analysis of its behavioral correctness.
The work is summarized and further research are discussed in Section~\ref{sec:conclusion}.

\section{Towards a methodology}
\label{sec:methodology}

\begin{figure*}[htb]
\centering
\begin{pspicture}(11.0,6.5) 
\pspolygon[linecolor=cyan,fillcolor=cyan,fillstyle=solid](2,0)(5,0)(3.5,6.2)

\psset{fillstyle=solid,fillcolor=gray,linecolor=gray,shadow=false}
\rput(3.5,5.8){\rnode{s1}{\psframebox{\textcolor{white}{\textbf{\textsc{1.~Use case diagrams}}}}}}
\rput(3.5,4.8){\rnode{s2}{\psframebox{\textcolor{white}{\textbf{\textsc{2.~Use case scenarios}}}}}}
\rput(8.5,4.8){$a_{1}, a_{2}, a_{3}, ...$}
\rput(3.5,3.8){\rnode{s3}{\psframebox{\textcolor{white}{\textbf{\textsc{3.~Activity diagrams}}}}}}
\rput(8.5,3.8){$P, W_{L}$}
\rput(3.5,2.8){\rnode{s4}{\psframebox{\textcolor{white}{\textbf{\textsc{4.~Logical spec.\ generation}}}}}}
\rput(8.5,2.8){${\cal A}(P,W_{L}) \longrightarrow L$}
\rput(3.5,1.8){\rnode{s5}{\psframebox{\textcolor{white}{\textbf{\textsc{5.~Defining properties}}}}}}
\rput(8.5,1.8){$Q$}
\rput(3.5,0.8){\rnode{s6}{\psframebox{\textcolor{white}{\textbf{\textsc{6.~Reasoning process}}}}}}
\rput(8.5,0.8){$C(L) \imp Q$}
\psset{fillstyle=none,linewidth=1.5pt}

\ncline[angleA=0,angleB=0]{->}{s1}{s2}
\ncline[angleA=0,angleB=0]{->}{s2}{s3}
\ncline[angleA=0,angleB=0]{->}{s3}{s4}
\ncline[angleA=0,angleB=0]{->}{s4}{s5}
\ncline[angleA=0,angleB=0]{->}{s5}{s6}
\nccurve[angleA=180,angleB=180]{->}{s6}{s5}

\rput(.6,5.8){\rnode{s10}{}}
\rput(.6,0.8){\rnode{s11}{}}
\ncline[linewidth=2pt]{*->}{s10}{s11}\naput[labelsep=-13pt,nrot=:D]{\textsf{Requirements model maturity}}
\rput(10.5,5.8){\rnode{s10}{}}
\rput(10.5,0.8){\rnode{s11}{}}
\ncline[linewidth=2pt]{*->}{s10}{s11}\nbput[labelsep=-13pt,nrot=:D]{\textsf{Logical modeling \& reasoning}}
\rput(3.5,.2){\rnode{s6}{\textcolor{white}{\textbf{\textsc{Formalization}}}}}
\end{pspicture}
\caption{Software requirements modeling and deduction-based verification}
\label{fig:formalization}
\end{figure*}
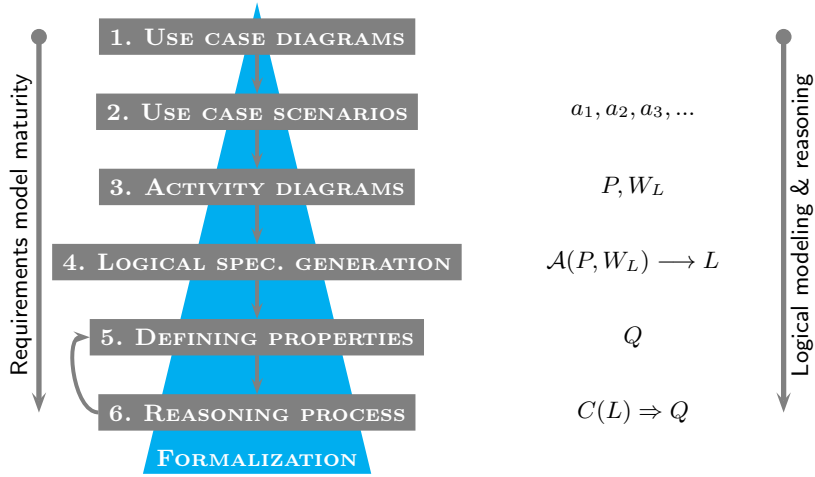

The outline of the procedure and guidelines used for the construction of a requirements model,
as it is understood in the work,
is briefly discussed below.
It constitutes a kind of methodology and its subsequent steps are presented in Fig.~\ref{fig:formalization}.
The entire procedure can be collected in the following items:
\begin{enumerate}
\item use case diagrams -- use case modeling to understand functions and goals of a system;
\item use case scenarios -- identifying and extracting atomic activities;
\item activity diagrams -- modeling workflows using predefined patterns;
\item automatic generation of logical specifications from requirements models;
\item manual definition of the desired model  properties;
\item formal verification of a desired property using the semantic tableaux method.
\end{enumerate}
All steps are shown on the left side of Fig.~\ref{fig:formalization}.
The first three steps involve the requirements modeling phase
but the last three steps involve generation of logical specification and analysis of
requirements model properties.
The loop between the last two steps
refers to a process of both identifying and verifying more and more new properties of
the examined model.
Some symbols and notation resulting from the introduced formalization are
on the right side of Fig.~\ref{fig:formalization} and
they are discussed in further sections of the work.
Generally, it leads, step by step, from an abstract view of a system
to more and more detailed and reliable and, finally, verified requirements models.

Let us summarize the entire method proposed in the work through
a se\-quence of the following steps:
{\small
\quad\rnode{pkt:ucd}{\ovalbox{$UCD_{i=1, ...}$}}
\qquad\quad\rnode{pkt:uc}{\ovalbox{$UC_{i=1, ...}$}}
\qquad\quad\rnode{pkt:a}{\ovalbox{$\{a_{1}, a_{2}, ...\}$}}\vspace{1.1\baselineskip}\\
\hspace*{.5cm}\rnode{pkt:ad}{\ovalbox{$AD_{i=1, ...}$}}
\qquad\quad\rnode{pkt:wl}{\ovalbox{$W_{L,i=1, ...}$}}
\qquad
\begin{minipage}{.2\linewidth}
\centering
\rnode{pkt:l}{\ovalbox{$L_{i=1, ...}$}}\\
\rnode{pkt:q}{\ovalbox{\quad$Q$\quad}}
\end{minipage}
\qquad\rnode{pkt:yn}{\ovalbox{$Y/N$}}
\ncline{->}{pkt:ucd}{pkt:uc}\Bput{\small $m$}
\ncline{->}{pkt:uc}{pkt:a}\Bput{\small $m$}
\ncangles[angleA=270,angleB=180,linearc=.15]{->}{pkt:a}{pkt:ad}\mput*{\small $m$}
\ncline{->}{pkt:ad}{pkt:wl}\Bput{\small $a$}
\ncline{->}{pkt:wl}{pkt:l}\Bput{\small $a$}
\ncline{->}{pkt:l}{pkt:yn}\Aput{\small $a$}
\ncline{->}{pkt:q}{pkt:yn}\Bput{\small $m$}
\quad
}
producing the polar answer $Y/N$,
where $m$ and $a$ mean manual or automatic, respectively, transition between steps,
and the answer is a reaction to a particular correctness question.
All these concepts and introduced notation are to be explained at the remainder of the work.

\section{Use cases and identification of activities}
\label{sec:use-cases-scenarios}

Defining use cases and scenarios is important not only to understand
 the functionalities of a system  but also to identify elementary activities.
The activities play an important role when building logical specifications,
i.e.\ the logical specification is modeled over atomic activities.
The \emph{use case diagram} consists of actors and use cases.
\emph{Actors} are objects which interact with a~system
and create the system's environment, thus providing interaction with the system.
\emph{Use cases} are services and functionalities which are used by actors.
The use case diagrams are a~rather descriptive technique and
do not refer to any details of the system implementation~\cite{Schneider-Winters-2001}.

\begin{figure}[htb]
\centering
\includegraphics[width=9.5cm]{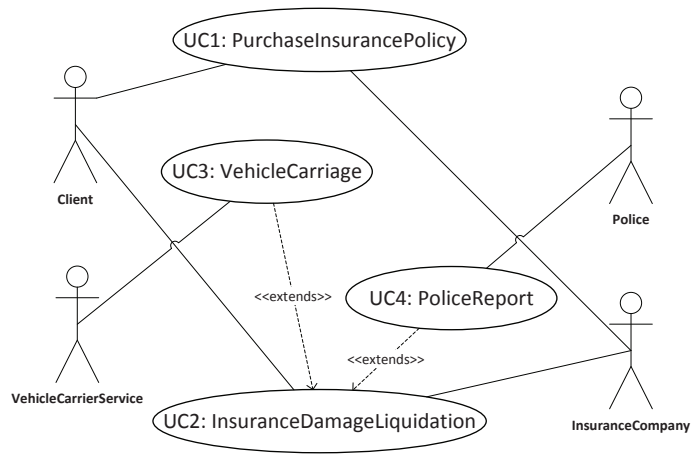}
\caption{A sample use case diagram $UCD$ ``CarInsuranceLiquidatingDamages''}
\label{fig:use-case-diagram}
\end{figure}
Let us present it more formally.
In the initial phase of a system modeling,
use case diagrams $UCD_{1}, ..., UCD_{n}$ are built.
Every $UCD_{i}$ diagram contains some use cases $UC_{1}, ..., UC_{m}$
which describe the desired functionality of a system.
A typical and sample use case diagram is shown in Fig.~\ref{fig:use-case-diagram}.
It consists of four actors and four use cases, $UC_{1}$, $UC_{2}$, $UC_{3}$ and $UC_{4}$,
modeling a system of car insurance and damages liquidation.
The diagram seems to be intuitive and is not discussed in detail.

Use cases are commonly used for capturing requirements through
\emph{scenarios} which are
brief narratives that describe the expected use of a system.
A scenario is a possible sequence of steps
which enables the achievement of a particular goal resulting from
the  functionality of a use case.
Every use case $UC_{i}$ has its own scenario.
From the point of view of the approach presented here,
scenarios play an additional important role, which is identification of atomic activities
used to build individual scenario steps.
An \emph{activity} is the smallest unit of computation.
Thus, every scenario contains some activities $a_{1}, ..., a_{n}$.
The set of \emph{atomic activities} $AA$ contains all activities
identified and defined for all scenarios.
The most valuable situation is when the entire use case scenario
involves identified activities and the narrative does not dominate
and is limited to model behavior, which is later formally shown in
activity diagrams.

\begin{figure}[htb]
\begin{tabular}{ll}
\begin{minipage}{.5\linewidth}
\centering
\includegraphics[width=6.0cm]{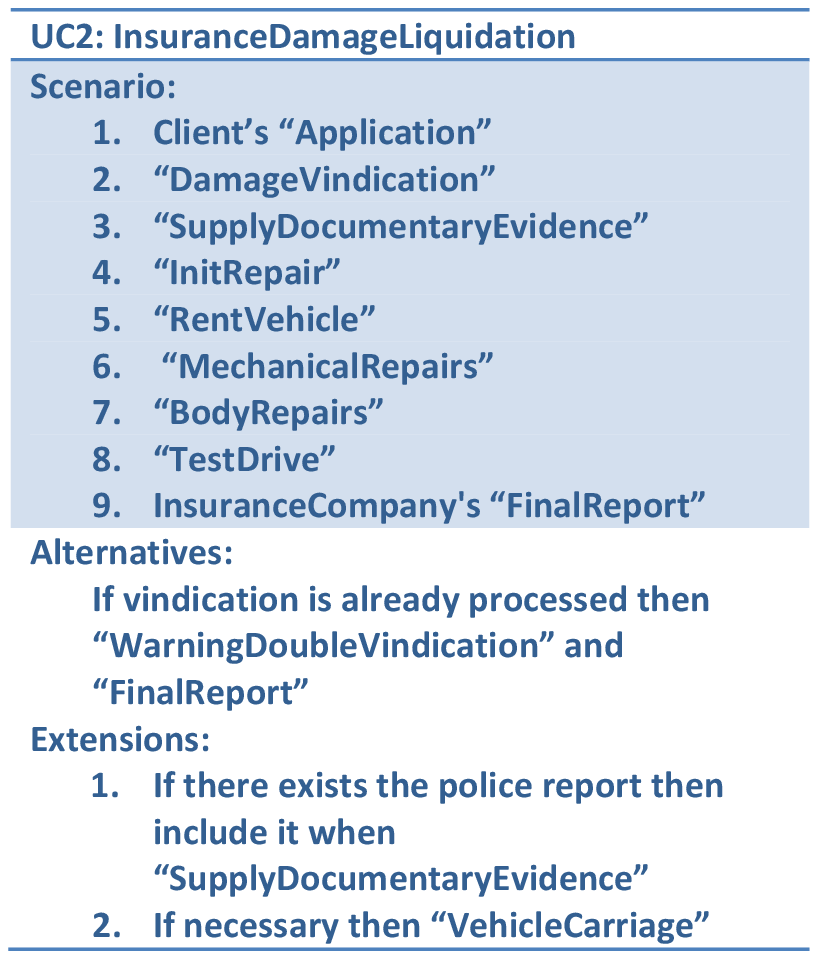}
\end{minipage}
&
\begin{minipage}{.5\linewidth}
\centering
\includegraphics[width=6.0cm]{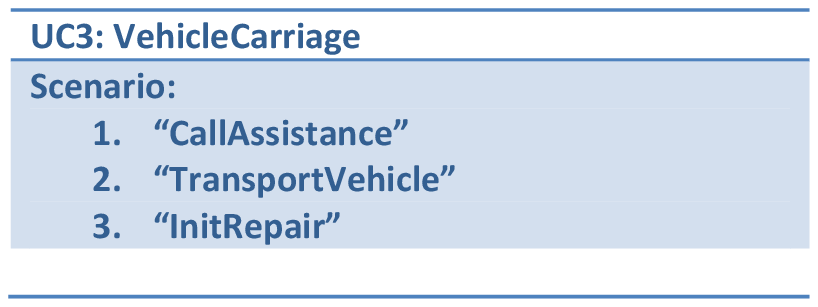}
\end{minipage}
\end{tabular}
\caption{Scenarios for use cases $UC_{2}$~``InsuranceDamageLiquidation'' (left) and extended $UC_{3}$~``VehicleCarriage'' (right)}
\label{fig:use-case-scenario}
\end{figure}
Sample scenarios for use cases $UC_{2}$ and $UC_{3}$,
i.e.\ ``InsuranceDamageLiquidation'' and ``VehicleCarriage'',
are shown in Fig.~\ref{fig:use-case-scenario}.
They contain some atomic activities which are identified
when preparing scenarios.
The alternative and extension points are defined.
The ``Application'' activity represents the act of applying for
an initiation of the compensation procedure.
``DamageVindication'' represents
the registration process in the insurer system,
verification of insurance,
and the start, if justified, of the process of recovery damages.
If all documentation is gathered (``SupplyDocumentaryEvidence''), it can start repairing (``InitRepair'').
If necessary then the use case is extended by the ``VehicleCarriage'' use case.
After the case registration and checking the insurance scope at the ``VehicleCarrierService'' actor,
the car is transported.
Let us note that the ``InitRepair'' activity is common to $UC_{2}$ and $UC_{3}$,
which shows a joint point and does not mean that the activity is executed more than once.
While the car repair process is carried out (``MechanicalRepairs'' and ``BodyRepairs''),
the client can hire a replacement vehicle (``RentVehicle'').
At the and of the scenario,
it is always generated a report (``FinalReport'').
The level of formalization presented here,
i.e.\ when discussing use cases and their scenarios,
is intentionally not very high.
This assumption seems realistic since this is
an initial phase of requirements modeling.
Dynamic aspects of activities are to be modeled strictly when
developing activity diagrams,
c.f.\ Section~\ref{sec:modeling-activities}.

\section{Logical background}
\label{sec:logical-background}

\emph{Temporal logic} TL introduces symbolism for reasoning about
truth and falsity of formulas throughout the flow of time,
taking the changes of their valuations into consideration.
Two basic and unary operators are $\som$ for ``sometime (or eventually) in the future''
and $\alw$ for ``always in the future''; these are dual operators.
Temporal logic exists in many varieties; however, these considerations are limited
to the \emph{linear-time temporal logic} or \emph{linear temporal logic} LTL.
Linear temporal logic refers to infinite sequences of computations
and attention is focused on the \emph{propositional linear time logic} PLTL.
These sequences refer to the Kripke structure which defines the semantics of TL,
i.e.\ a syntactically correct formula can be satisfied by
an infinite sequence of truth evaluations over a set of \emph{atomic propositions} AP.
It should be pointed out that atomic propositions are identical to atomic activities defined
in Section~\ref{sec:use-cases-scenarios},
i.e.\ $AA=AP$.
The basic issues related to temporal logics and their syntax and semantics are discussed
in many works, e.g.~\cite{Emerson-1990,Wolter-Wooldridge-2011}.

The properties of the time structure are fundamental to a logic.
Of particular significance is
the \emph{smallest\emph{, or \emph{minimal},} temporal logic}, e.g.~\cite{Chellas-1980,vanBenthem-1995},
also known as temporal logic of class~K.
The minimal temporal logic is an extension of the classical propositional calculus of the axiom
$\alw (\Phi \imp \Psi) \imp (\alw \Phi \imp \alw \Psi)$
and the inference rule $\theo \Phi \Longrightarrow \theo \alw \Phi$.
The essence of the logic is the fact that there are
no specific assumptions for the properties of the time structure.
The following formulas may be considered as typical examples:
$action \imp \som reaction$, $\alw(send \imp\som receive)$, $\som alive$, $\alw\neg (badevent)$, etc.
The considerations of the work are limited to this logic since
it allows to define many system properties (safety, liveness);
it is also easier to build a deduction engine,
or use  existing verified provers,
and to  quickly verify the approach proposed in the work.

Although the work is not based on any particular method of reasoning,
the method of semantic tableaux is presented in a more detailed way.
\emph{Semantic tableaux} is a decision-making procedure
for checking satisfiability of a formula.
The method is well known in classical propositional logic but it can also be
applied in modal and temporal logics~\cite{Agostino-etal-1999},
and for the propositional linear-time logic first presented in~\cite{Wolper-1985}.
The method is based on formula decompositions.
In the semantic tableaux method,
at the end of the decomposition procedure, all branches of the
received tree are searched for contradictions.
When all branches of a tree have contradictions,
it means that the inference tree is \emph{closed}.
If a negation of the initial formula is placed in the root,
this leads to the statement that the initial formula is true.
This method has some advantages over the traditional axiomatic approach.
In the classical reasoning approach,
starting from axioms, longer and more complicated
formulas are generated and derived.
Formulas become longer and longer step by step,
and only one of them will lead to the verified formula.
The method of semantic tableaux is characterized
by a reverse strategy.
The method provides, through so-called \emph{open}
branches of the semantic tree, information about the source of an
error, if one is found, which is another and very important
advantage of the method.
Summing up, the tableaux are global, goal-oriented and ``backward'',
while resolution is local and ``forward''.

\begin{figure}[htb]
\centering
{\small
\pstree[levelsep=4.5ex,nodesep=2pt,treesep=20pt]
       {\TR{$[1,-]1: \neg((\alw(a \imp \som b) \con \alw(b \imp \som c)) \imp \alw(a \imp \som c))$}}{
          \pstree{\TR{$[2,1]1: \alw(a \imp \som b) \con \alw(b \imp \som c) \con \som a \con \alw\neg c$}}{
            \pstree{\TR{$[3,2]1: \som a$}}{
              \pstree{\TR{$[4,2]1: \alw\neg c$}}{
                \pstree{\TR{$[5,2]1: \alw(b \imp \som c)$}}{
                  \pstree{\TR{$[6,2]1: \alw(a \imp \som b)$}}{
                    \pstree{\TR{$[7,3]1.[a]: a$}}{
                      \pstree{\TR{$[8,4]1.(x): \neg c$}}{{
                           \pstree{\TR{$[9,5]1.(y): b \imp \som c$}}{{
                               \pstree{\TR{$[10,9]1: \neg b$}}{
                                 \pstree{\TR{$[13,6]1.(z): a \imp \som b$}}{{
                                     \pstree{\TR{$[15,13]1: \neg a$}}{$\times$}
                                     \pstree{\TR{$[16,13]1: \som b$}}{
                                       \pstree{\TR{$[17,16]1.[b]: b$}}{$\times$}}}}}
                               \pstree{\TR{$[11,9]1: \som c$}}{
                                 \pstree{\TR{$[12,11]1.[c]: c$}}{
                                   \pstree{\TR{$[14,6]1.[z]: a \imp \som b$}}{{
                                       \pstree{\TR{$[18,14]1: \neg a$}}{$\times$}
                                       \pstree{\TR{$[19,14]1: \som b$}}{
                                         \pstree{\TR{$[20,19]1.[b]: b$}}{$\times$}}}}}}}}}}}}}}}}}
}
\caption{A sample inference tree}
\label{fig:inference-tree}
\end{figure}
A simple yet illustrative example of an inference tree is shown in the left side of Fig.~\ref{fig:inference-tree}.
The relatively short formula gives a small inference tree,
but shows how the method works.
The label $[i,j]$ means that it is the $i$-th formula, i.e.\ the $i$-th decomposition step,
received from the decomposition transformation of a formula stored in the $j$-th node.
The label ``$1:$'' represents the initial world in which a formula is true.
The label ``$1.(x)$'', where $x$ is a free variable,
represents all possible worlds that are consequences of world~$1$.
On the other hand, the label ``$1.[p]$'',
where $p$ is an atomic formula,
represents one of the possible worlds,
i.e.\ a successor of world~$1$,
where formula $p$ is true.
The decomposition procedure adopted and presented here
refers to the first-order predicate calculus
and can be found, for example, in the work~\cite{Hahnle-1998}.
All branches of the analyzed trees are closed ($\times$).
There is no valuation that satisfies the root formula.
This, consequently, means that
the formula before the negation is always satisfied.

\begin{figure}[htb]
\centering
\begin{pspicture}(5,6.5) 
\psset{framearc=0}
\psset{shadow=true,shadowcolor=gray}
\rput(2.5,5.8){\rnode{r1}{\psframebox
                      {\begin{tabular}{c}
                            \textcolor{black}{\textsc{UML}}\\
                            \textcolor{black}{\textsc{Modeler}}
                      \end{tabular}}}}
\rput(2.5,3.6){\rnode{r2}{\psframebox
                      {\begin{tabular}{c}
                            \textcolor{black}{\textsc{Logical Specifica-}}\\
                            \textcolor{black}{\textsc{tions Generator} \psframebox[boxsep=true,shadow=false]{\scriptsize G}}
                      \end{tabular}}}}
\rput(0.8,2.3){\rnode{r3}{$P$}}
\rput(2.5,1.5){\rnode{r8}{\psframebox
                      {\begin{tabular}{c}
                            \textcolor{black}{\textsc{ST Temporal}}\\
                            \textcolor{black}{\textsc{Prover} \psframebox[linewidth=1pt,boxsep=true,shadow=false]{\scriptsize T}}
                      \end{tabular}}}}
\rput(0,1.5){\rnode{r7}{$Q$}}
\rput(2.5,0.0){\rnode{r10}{Y/N}}

\psset{linewidth=1.5pt}
\psset{shadow=false}
\psset{linecolor=black}
\ncline{->}{r1}{r2}\Bput{Requir.}\Aput{models}
\ncline[nodesep=.08cm]{->}{r3}{r2}
\ncline{->}{r2}{r8}\Aput{$L$}
\ncline[nodesep=.08cm]{->}{r7}{r8}
\ncline{->}{r8}{r10}
\psframe[linestyle=dashed,linewidth=1pt](0.5,0.6)(4.6,4.5)
\end{pspicture}
\caption{A deduction-based verification system}
\label{fig:deduction-system}
\end{figure}
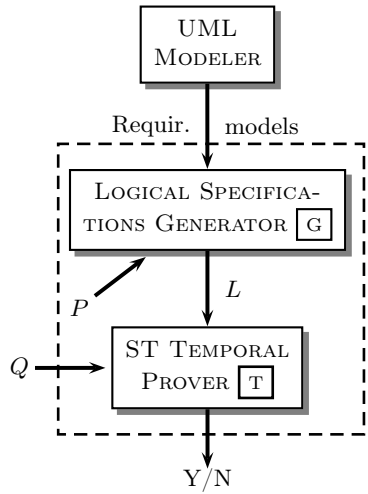
An outline architecture of the proposed
deduction-based verification system is presented in Fig.~\ref{fig:deduction-system}.
A similar system is proposed in work~\cite{Klimek-2013-kes-amsta}.
The system works automatically and consists of some important elements.
The \psframebox[boxsep=true,shadow=false]{\scriptsize G} component
generates logical specifications which are
sets of a usually large number of temporal logic formulas (of class K).
Formula generation is performed automatically from
workflow models, which are constructed from predefined patterns for activity diagrams.
The extraction process is discussed in Section~\ref{sec:generating-logical-specifications}.
The whole specification $L$ can be treated as a~conjunction of formulas
$f_{1}\con \ldots \con f_{n} = C(L)$,
where every $f_{i}$ is a formula generated during the extraction process.
The $Q$ formula is a desired property for a requirements model.
Both the system specification and the examined properties are input to
the \psframebox[boxsep=true,shadow=false]{\scriptsize T} component,
i.e.\ \emph{Semantic Tableaux Temporal Prover},
or shortly \emph{ST Temporal Prover}, which enables
the automated reasoning in temporal logic using semantic tableaux.
The input for this component is the formula $C(L) \imp Q$, or, more precisely:
\begin{eqnarray}
f_{1}\con \ldots \con f_{n} \imp Q \label{initial-formula}
\end{eqnarray}
Due to the fact that the semantic tableaux method is an indirect proof,
i.e.\ \emph{reductio ad absurdum},
 after the negation of Formula~\ref{initial-formula},
it is placed at the root of the inference tree
and decomposed using well-defined rules of the semantic tableaux method.
If the inference tree is closed, it means that
the initial Formula~\ref{initial-formula} is true.
The output of the \psframebox[boxsep=true,shadow=false]{\scriptsize T} component,
and therefore also the output of the entire deductive system, is the answer Yes/No.
This output also realizes the final step of the procedure shown in Fig.~\ref{fig:formalization}.
However,
the verification procedure can be performed for the further properties,
c.f.\ the loop in Fig.~\ref{fig:formalization}.

The verification procedure which results from the deduction system
in Fig.~\ref{fig:deduction-system} can be summarized as follows:
\begin{enumerate}
\item automatic generation of system specifications
      (the \psframebox[shadow=false,boxsep=true]{\scriptsize G} component);
\item introduction of the property $Q$ of the system;
\item automatic inference using semantic tableaux
      (the \psframebox[boxsep=true,shadow=false]{\scriptsize T} component)
      for the whole complex formula, c.f.\ Formula~\ref{initial-formula}.
\end{enumerate}
Steps~1 to~3, in whole or individually, may be processed many times,
whenever the specification of the UML model is changed (step~1) or
if there is a need for a new inference due to
the revised system specification (steps~2 or~3).

The prover is an important component of the architecture for the deduction-based system shown in Fig.~\ref{fig:deduction-system}.
It enables automate the inferencing process and formal verification of developed models.
Reasoning engines are more available,
especially in recent years when a number of provers for modal logics become accessible,
c.f.\ Schmidt~\cite{Schmidt-2013-provers}.
Work by Gor{\'e} et al.~\cite{Gore-2011}
provides experimental results for some existing provers,
which are based on different methods of reasoning.
Another work by Hustadt and Schmidt~\cite{Hustadt-Schmidt-1999}
provides also an experimental performance analysis of
some other theorem provers based on different
method of reasoning, and some randomly generated formulas are analysed.
Work by Islam et al.~\cite{Islam-etal-2012} provides
a brief overview of some existing tableau theorem provers.
In work by Dixon et al.~\cite{Dixon-etal-2012}
an automated tableau prover generator is used and some implementation and experimental results are discussed.
Summing up, selection of an appropriate existing prover,
or building one's own,
constitutes a separate task that exceeds the size and main objectives of the work,
c.f.\ also the concluding remarks in the last Section~\ref{sec:conclusion}.

\section{Workflow patterns and modeling activities}
\label{sec:modeling-activities}

Activity diagrams constitute a closure of the development phase for requirements models,
by introducing dynamic aspects for models.
This aspect is subjected to the correctness analysis for safety and liveness properties.
The \emph{activity diagram} enables modeling of workflow activities.
It constitutes a graphical representation of workflow
showing the flow of control from one activity to another.
It supports choice, concurrency and iteration.
The \emph{swimlane} is useful for partitioning the activity diagram
and enables grouping of activities in a single thread.
The important goal of activity diagrams is to show how an activity depends on others~\cite{Pender-2003}.

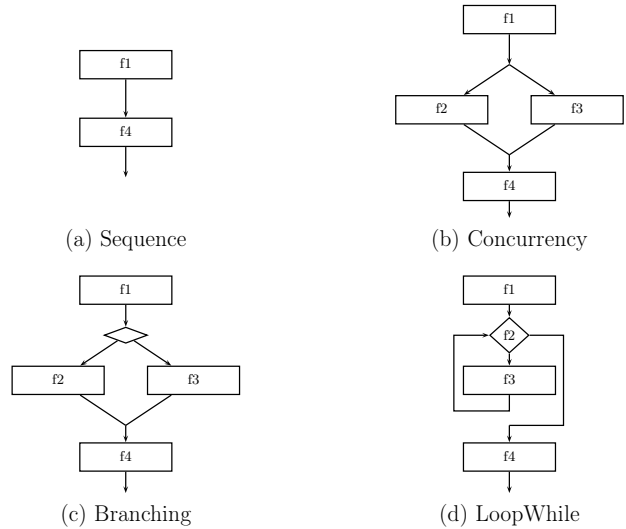
\begin{figure}[htb]
\centering
\begin{pspicture}(8.2,7.3) 
      \psset{linecolor=black}
      \scalebox{0.6}{
      \rput(2.5,10.5){\rnode{p1-1}{\psframebox{\makebox[1.8cm][c]{\begin{tabular}{c}\mbox{f1}\\\end{tabular}}}}}
      \rput(2.5,9.0){\rnode{p1-2}{\psframebox{\makebox[1.8cm][c]{\begin{tabular}{c}\mbox{f4}\\\end{tabular}}}}}
      \rput(2.5,8.0){\pnode{p1-3}{}}
      \ncline{->}{p1-1}{p1-2}
      \ncline{->}{p1-2}{p1-3}
      \rput(2.5,6.6){\Large (a)~Sequence}

      \rput(11.0,11.5){\rnode{p2-1}{\psframebox{\makebox[1.8cm][c]{\begin{tabular}{c}\mbox{f1}\\\end{tabular}}}}}
      \rput(11.0,10.5){\pnode{p2-1b}{}}
      \rput(9.5,9.5){\rnode{p2-2}{\psframebox{\makebox[1.8cm][c]{\begin{tabular}{c}\mbox{f2}\\\end{tabular}}}}}
      \rput(12.5,9.5){\rnode{p2-3}{\psframebox{\makebox[1.8cm][c]{\begin{tabular}{c}\mbox{f3}\\\end{tabular}}}}}
      \rput(11.0,7.8){\rnode{p2-4}{\psframebox{\makebox[1.8cm][c]{\begin{tabular}{c}\mbox{f4}\\\end{tabular}}}}}
      \rput(11.0,8.5){\pnode{p2-5}{}}
      \rput(11.0,7.1){\pnode{p2-6}{}}
      \ncline{->}{p2-1}{p2-1b}
      \ncline{->}{p2-1b}{p2-2}
      \ncline{->}{p2-1b}{p2-3}
      \ncline{-}{p2-2}{p2-5}
      \ncline{-}{p2-3}{p2-5}
      \ncline{->}{p2-5}{p2-4}
      \ncline{->}{p2-4}{p2-6}
      \rput(11.0,6.6){\Large (b)~Concurrency}

      \rput(2.5,5.5){\rnode{p3-1}{\psframebox{\makebox[1.8cm][c]{\begin{tabular}{c}\mbox{f1}\\\end{tabular}}}}}
      \rput(2.5,4.5){\dianode{p3-1b}{\quad}}
      \rput(1,3.5){\rnode{p3-2}{\psframebox{\makebox[1.8cm][c]{\begin{tabular}{c}\mbox{f2}\\\end{tabular}}}}}
      \rput(4,3.5){\rnode{p3-3}{\psframebox{\makebox[1.8cm][c]{\begin{tabular}{c}\mbox{f3}\\\end{tabular}}}}}
      \rput(2.5,2.5){\pnode{p3-3b}{}}
      \rput(2.5,1.8){\rnode{p3-4}{\psframebox{\makebox[1.8cm][c]{\begin{tabular}{c}\mbox{f4}\\\end{tabular}}}}}
      \rput(2.5,1.0){\pnode{p3-6}{}}
      \ncline{->}{p3-1}{p3-1b}
      \ncline{->}{p3-1b}{p3-2}
      \ncline{->}{p3-1b}{p3-3}
      \ncline{-}{p3-2}{p3-3b}
      \ncline{-}{p3-3}{p3-3b}
      \ncline{->}{p3-3b}{p3-4}
      \ncline{->}{p3-4}{p3-6}
      \rput(2.5,0.5){\Large (c)~Branching}

      \rput(11.0,5.5){\rnode{p4-1}{\psframebox{\makebox[1.8cm][c]{\begin{tabular}{c}\mbox{f1}\\\end{tabular}}}}}
      \rput(11.0,4.5){\dianode{p4-2}{f2}}
      \rput(11.0,3.5){\rnode{p4-3}{\psframebox{\makebox[1.8cm][c]{\begin{tabular}{c}\mbox{f3}\\\end{tabular}}}}}
      \rput(11.0,2.5){\pnode{p4-5}{}}
      \rput(11.0,2.0){\pnode{p4-6}{}}
      \rput(11.0,1.8){\rnode{p4-4}{\psframebox{\makebox[1.8cm][c]{\begin{tabular}{c}\mbox{f4}\\\end{tabular}}}}}
      \rput(11.0,1.0){\pnode{p4-7}{}}
      \ncline{->}{p4-1}{p4-2}
      \ncline{->}{p4-2}{p4-3}
      \ncloop[angleA=270,angleB=180,loopsize=35pt]{->}{p4-3}{p4-2}
      \ncangle[armB=1.2cm]{-}{p4-2}{p4-5}
      \ncline{->}{p4-5}{p4-4}
      \ncline{->}{p4-4}{p4-7}
      \rput(11.0,0.5){\Large (d)~LoopWhile}
       }
\end{pspicture}
\caption{Workflow patterns for activities}
\label{fig:activity-workflow}
\end{figure}
From the viewpoint of the approach presented in the work,
it is important to introduce a number of predefined workflow patterns for
activities that provide all workflows in a structural form.
A \emph{pattern} is a generic description of the structure of some computations.
Nesting of patterns is permitted.
The following workflow patterns are predefined:
\emph{sequence},
\emph{concurrent fork/join},
\emph{branching}
and \emph{loop while} for iteration as
they are shown in Fig.~\ref{fig:activity-workflow}.
It is assumed that only predefined patterns can be used for
modeling  of activity  behavior.
Such structuring is not a limitation when modeling arbitrarily complex sets of activities.

\begin{figure}[htb]
\begin{tabular}{ll}
\begin{minipage}{.6\linewidth}
\centering
\includegraphics[width=6.5cm]{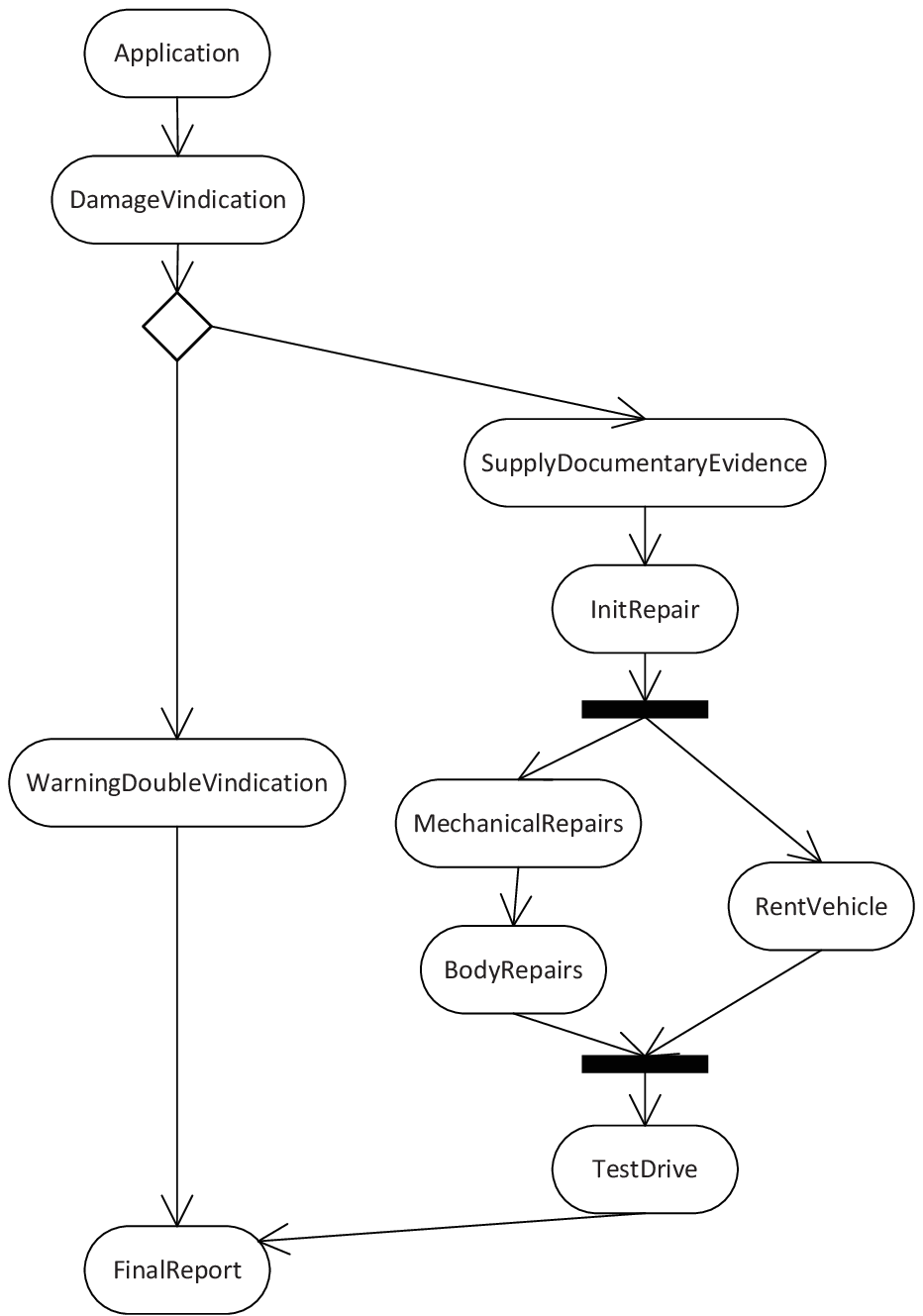}
\end{minipage}
&
\begin{minipage}{.3\linewidth}
\centering
\includegraphics[width=3.0cm]{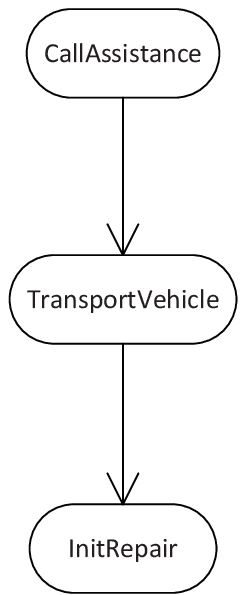}
\end{minipage}
\end{tabular}
\caption{Sample activity diagrams $AD_{2}$ (left) and $AD_{3}$ (right) for use cases $UC_{2}$~``InsuranceDamageLiquidation'' and $UC_{3}$~``VehicleCarriage''}
\label{fig:activity-diagram}
\end{figure}
For every use case $UC_{i}$ and its scenario,
a activity diagram $AD_{i}$ is developed/modeled.
The activity diagram workflow is modeled only using
atomic activities which are
identified when building a use case scenario.
Furthermore, workflows are composed only using the predefined design patterns
shown in Fig.~\ref{fig:activity-workflow}.
Sample activity diagrams $AD_{2}$ and $AD_{3}$ are shown in Fig.~\ref{fig:activity-diagram}.
They model behavior of the $UC_{2}$ and $UC_{3}$ use cases shown in Fig.~\ref{fig:use-case-diagram},
using activities from the scenarios in Fig.~\ref{fig:use-case-scenario}.
After the start of the vindication process, i.e.\ ``DamageVindication'',
it is checked whether it is already being processed.
If yes, the decision to register this fact is made, as it is likely another attempt at vindication of the same event,
c.f.\ ``WarningDoubleVindication''.
The scenario analysis and the nature of other activities,
i.e.\ ``MechanicalRepairs'', ``BodyRepairs'' and ``RentVehicle'',
leads to the conclusion that they can and should be performed concurrently.

This step completes the phase of modeling requirements, c.f.\ Fig.~\ref{fig:formalization}.
The next steps involve the widely understood formal analysis of obtained requirements models.

\section{Generating logical specifications}
\label{sec:generating-logical-specifications}

The phase of modeling requirements is complete when all activity diagrams
for all scenarios are built, c.f.\ Fig.~\ref{fig:formalization} and Section~\ref{sec:modeling-activities}.
Then, the phase of generating logical specifications and formal analysis of the desired properties begins.
The logical specification generation process must be performed in an automatic way.
Such logical specifications usually consist of a large number of temporal logic formulas
and their manual development is practically impossible since
this process can be monotonous, error-prone
and the creation of such logical specifications
is difficult for inexperienced analysts.
On the other hand, the verified properties of the system constitute usually easier formulas,
not to mention the fact that they are rather individual temporal logic formulas.

The proposed algorithm for automatic extraction of
logical specifications is based on the assumption that
all workflows for activity diagrams are built using
only well-known workflow patterns, c.f.\ Fig.~\ref{fig:activity-workflow}.
The process of building a~logical specification can be presented in the following steps:
\begin{enumerate}
\item analysis of activity diagrams to extract all predefined workflow patterns,
\item translation of the extracted patterns to a~logical expression $W_{L}$,
\item generating a~logical specification $L$ from logical expressions,
      i.e.\ receiving a set of temporal logic formulas.
\end{enumerate}

Predefined workflow patterns constitute a kind of primitives which are defined using
temporal logic formulas. Therefore,
an \emph{elementary set} $pat()$ of formulas over atomic formulas $a_{i}$,
where $i>0$, which is also denoted $pat(a_{i})$,
is a set of temporal logic formulas $f_{1}, ..., f_{m}$
such that all formulas are syntactically correct (and restricted to the logic~K).
For example, an elementary set
$pat(a,b,c,d)=\{a\imp\som b, b \imp \som(c \dis d), \alw\neg((a \dis b) \con\neg c)\}$ is
a three-element set of PLTL formulas, created over four atomic formulas.
Let $\Sigma$ be a set of \emph{predefined design patterns},
i.e.\ $\Sigma=\{ Sequence,\ Concurrency,\ Branching,\ LoopWhile \}$.
The proposed temporal logic formulas should describe
both safety and liveness properties of each pattern.
Let us introduce some aliases:
$Seq$ as $Sequence$, $Concur$ as $Concurrency$,
$Branch$ as $Branching$ and $Loop$ as $LoopWhile$.

Every activity workflow is designed using only predefined design patterns.
Every design pattern has a~predefined and countable set of linear temporal logic formulas.
The workflow model can be quite complex and it may contain nesting patterns.
Let us define a logical expression, which is similar to well known regular expressions,
to represent any potentially complex structure of the activity workflow
but also to have a literal representation for these workflows.
The \emph{logical expression} $W_{L}$ is a structure created using the following rules:
\begin{itemize}
\item every elementary set $pat(a_{i})$,
      where $i>0$ and every $a_{i}$ is an atomic formula,
      is a logical expression,
\item every $wrf(A_{i})$, where $i>0$ and every $A_{i}$ is either
      \begin{itemize}
      \item an atomic formula, or
      \item a logical expression $pat()$,
      \end{itemize}
      is also a logical expression.
\end{itemize}
Examples of logical expressions are given in Section~\ref{sec:reasoning-verification}.

Some restrictions on set of atomic formulas $a_{1}, \ldots, a_{n}$ of the logical expression $pat()$,
due to the number of arguments and their partial order, are introduced.
$a_{1}$ is always the first argument called an \emph{entry argument/formula},
and closely there is one entry argument.
$a_{4}$ is always the last argument called an \emph{exit argument/formula},
and closely there is one exit argument.
The subset of arguments between, informally speaking,
the first and the last argument are \emph{ordinary arguments} and this subset may be empty.
This implies that there are at least two arguments, but not more than four arguments.
From this it also follows that there is also only one entry, or exit, respectively,
point for a pattern, c.f.\ Fig~\ref{fig:predefined-P}.
The further discussion on these limitations, especially number of arguments $n=4$,
is provided in the end of Section~\ref{sec:generating-logical-specifications}.
Let us also note that entry and exit formulas enable representing the pattern
as a whole, i.e.\ without analysis of its internal behavior if it is necessary.

\begin{figure}[htb]
\centering
{\normalsize
\begin{tabular}{|l|}
\hline
\begin{minipage}{.9\textwidth}
\begin{verbatim}
                                     /* ver. 6.01.2014 */
Sequence(f1,f4):
f1 => <>f4 / ~f1 => ~<>f4 / []~(f1 & f4)
Concurrency(f1,f2,f3,f4):
f1 => <>f2 & <>f3 / ~f1 => ~(<>f2 & <>f3)
f2 & f3 => <>f4 / ~(f2 & f3) => ~<>f4
[]~(f1 & (f2 | f3)) / []~((f2 | f3) & f4) / []~(f1 & f4)
Branching(f1,f2,f3,f4):
f1 => (<>f2 & ~<>f3) | (~<>f2 & <>f3)
~f1 => ~((<>f2 & ~<>f3) | (~<>f2 & <>f3))
f2 | f3 => <>f4 / ~(f2 | f3) => ~<>f4
[]~(f1 & f4) / []~(f2 & f3)
[]~(f1 & (f2 | f3)) / []~((f2 | f3) & f4)
LoopWhile(f1,f2,f3,f4):
f1 => <>f2 / ~f1 => ~<>f2
f2 & c(f2) => <>f3 & ~<>f4
~(f2 & c(f2)) => ~(<>f3 & ~<>f4)
f2 & ~c(f2) => ~<>f3 & <>f4
~(f2 & ~c(f2)) => ~(~<>f3 & <>f4)
f3 => <>f2 / ~f3 => ~<>f2
[]~(f1 & f2) / []~(f1 & f3) / []~(f1 & f4)
[]~(f2 & f3) / []~(f2 & f4) / []~(f3 & f4)

\end{verbatim}
\end{minipage}\\
\hline
\end{tabular}
}
\caption{A predefined set of patterns $P$ and their temporal properties}
\label{fig:predefined-P}
\vspace{-1\baselineskip}
\end{figure}
The last step is to define a logical specification which is
generated from logical expressions.
The \emph{logical specification} $L$ consists of all formulas derived from
a logical expression $W_{L}$ using the algorithm ${\cal A}$,
i.e.\ $L(W_{L}) = \{f_{i} : i \geq 0 \con f_{i} \in {\cal A}(W_{L},P)\}$,
where $f_{i}$ is a TL formula.
Generating logical specifications is not a simple summation of formula collections
resulting from a logical expression.
The generation algorithm has two inputs.
The first one is a logical expression $W_{L}$ which is a kind of variable,
i.e.\ it varies for every (workflow) model, when the workflow is subjected to any modification.
The second one is a predefined set~$P$ which is a kind of constant,
i.e.\ once defined then widely used.
The example of such a set is shown in Fig~\ref{fig:predefined-P}.
All formulas describe both safety and liveness properties for a pattern~\cite{Alpern-Schneider-1985}.
However, the formulas are not discussed in the work because they might be
a subject of consideration in a separate work.
Moreover, the formulas can and should be prepared by an expert
with skills and theoretical background.
It guarantees that an inexperienced software analyst or engineer will be able to obtain correct logical models.
Most elements of the predefined $P$~set,
i.e.\ comments, two temporal logic operators, classical logic operators,
are not in doubt.
The slash allows to place more formulas in a single line.
$f_{1}$, $f_{2}$ etc.\ are atomic formulas for a pattern.
They constitute a kind of formal arguments for a pattern.
$\som f$ means that sometime (or eventually in the future), activity $f$ is satisfied,
i.e.\ the token reaches the activity.
$c(f)$ means that a logical condition associated with activity~$f$ has been evaluated and is satisfied.
For example, the expression $f \con c(f)$ means that the $f$ activity is satisfied (executed)
and $c(f)$ is true (neither false nor undetermined).
When the rising edge for the $f$ activity is reached,
the $c(f)$ becomes undetermined until it is evaluated to true or false.
When $c(f)$ is undetermined,
then the evaluation of the entire sample expression $f \con c(f)$ is
stopped until the evaluation is possible,
i.e.\ all elements of the expression are determined.

The notion of joined entry and exit formulas is introduced which is a result
of nested workflows, as well as the need to transfer,
informally speaking,
the logical signal to all start/termination points of a nested workflow.
Let $w$ is a logical expression,
then $w^{\blacktriangleright}$ is the \emph{joined entry formula},
and $w^{\blacktriangleleft}$ is the \emph{joined exit formula},
when the joined formula is calculated using the following (recursive) rules:
\begin{enumerate}
\item if there is no workflow itself in the place of the first atomic formula/argument
      (the $f_{1}$ formula), or the last atomic formula/argument (the $f_{4}$ formula), respectively,
      then $w^{\blacktriangleright}$ is equal to $f_{1}$, or $w^{\blacktriangleleft}$ is equal to $f_{4}$, respectively,
\item if there is a workflow, say $t()$, in a place of the first argument,
      i.e.\ the $f_{1}$ formula, or the last argument, i.e.\ the $f_{4}$ formula, respectively,
      then it is replaced by $t()^{\blacktriangleright}$, or $t()^{\blacktriangleleft}$, respectively for every such case.
\end{enumerate}
These rules allow to define joined point formulas for an arbitrary complex logical expression.
For example, for logical expression $w=Seq(a,b)$ formulas are $w^{\blacktriangleright}=a$ and $w^{\blacktriangleleft}=b$.
For expression $w=Branch(a,b,c,Seq(d,e))$ formulas are $w^{\blacktriangleright}=a$ and
$w^{\blacktriangleleft}=Seq(d,e)^{\blacktriangleleft}=e$
(by the way, for nested $Seq(d,e)^{\blacktriangleright}=d$).
For expression $w=Seq(Concur(Seq(a,b),c,d,e),f)$ formulas are
$w^{\blacktriangleright}=Concur(Seq(a,b),c,d,e)^{\blacktriangleright}=Seq(a,b)^{\blacktriangleright}=a$
and $w^{\blacktriangleleft}=f$.

\begin{algorithm}[htb]
\caption{Generating logical specifications (${\cal A}$)}
\label{alg:generating-specification}
{\normalsize
\begin{algorithmic}[1]
\algrenewcommand\algorithmicrequire{\textbf{Input:}}
\algrenewcommand\algorithmicensure{\textbf{Output:}}
\Require Logical expression $W_{L}$ (non-empty), predefined set $P$ (non-empty)
\Ensure Logical specification $L$
\State $L:=\emptyset$ \Comment{initiating specification}\label{alg:2:atomic-ini}
\For{every workflow $wrf()$ of $W_{L}$ from left to right}\label{alg:2:atomic-for}
\If{all arguments of $wrf()$ are atomic}\label{alg:2:atomic-s}
\State{$L := L \cup wrf()$}
\EndIf\label{alg:2:atomic-e}
\If{any argument of $wrf()$ is a workflow pattern itself}\label{alg:2:non-atomic-s}
\State{for every such an argument, say $r()$, substitute}
\State{disjunction of its joined entry and exit}
\State{formulas in all places where the argument}
\State{occurs in the $wrf()$ temporal formulas, i.e.}
\State{$L := L \cup (wrf() \leftarrow \mbox{``$r()^{\blacktriangleright} \dis r()^{\blacktriangleleft}$''})$}\label{alg:2:non-atomic-dis}
\EndIf\label{alg:2:non-atomic-e}
\EndFor
\end{algorithmic}
}
\end{algorithm}
\psset{linecolor=gray,linewidth=.6pt}
The output of the generation algorithm is a logical specification understood as
a set of temporal logic formulas.
The generation algorithm (${\cal A}$) is given as Algorithm~\ref{alg:generating-specification}.
Let us assume some auxiliary constraints for the Algorithm.
It is mandatory for every two patterns to have disjoint sets of atomic activities (arguments).
Every pattern consists of at least two activities/tasks (arguments),
c.f.\ Fig.~\ref{fig:activity-workflow} or Fig.~\ref{fig:predefined-P}.
Let us supplement the Algorithm by some examples.
The example for lines \ref{alg:2:atomic-s}--\ref{alg:2:atomic-e}:
$Concur(a,b,c,d)$ gives
$L=\{a \imp \som b \con \som c,
\neg a \imp \neg(\som b \con \som c),
b \con c \imp \som d,
\neg(b \con c) \imp \neg\som d,
\alw\neg(a \con (b \dis c)),
\alw\neg((b \dis c) \con d),
\alw\neg(a \con d)\}$.
The example for lines \ref{alg:2:non-atomic-s}--\ref{alg:2:non-atomic-e}:
$Branch(Seq(a,b),c,d,e)$ leads to
$L = \rnode{alg:non-atomic-a}{\ovalbox{\{}}
(a \dis b) \imp (\som c \con \neg\som d) \dis (\neg\som c \con \som d),
\neg(a \dis b) \imp \neg((\som c \con \neg\som d) \dis (\neg\som c \con \som d)),
c \dis d \imp \som e,
\neg(c \dis d) \imp \neg\som e,
\alw\neg((a \dis b) \con e),
\alw\neg(c \con d),
\alw\neg((a \dis b) \con (c \dis d)),
\alw\neg((c \dis d) \con e)
\}
\cup
\rnode{alg:atomic-a}{\ovalbox{\{}}
a \imp \som b,
\neg a \imp \neg\som b,
\alw\neg(a \con b)
\}
\rnode{alg:sum-set-a}{\ovalbox{=\{}}
(a \dis b) \imp (\som c \con \neg\som d) \dis (\neg\som c \con \som d),
\neg(a \dis b) \imp \neg((\som c \con \neg\som d) \dis (\neg\som c \con \som d)),
c \dis d \imp \som e,
\neg(c \dis d) \imp \neg\som e,
\alw\neg((a \dis b) \con e),
\alw\neg(c \con d),
\alw\neg((a \dis b) \con (c \dis d)),
\alw\neg((c \dis d) \con e),
a \imp \som b,
\neg a \imp \neg\som b,
\alw\neg(a \con b)
\}$.
The first set follows directly from lines
\rnode{alg:non-atomic-b}{\ovalbox{\ref{alg:2:non-atomic-s}--\ref{alg:2:non-atomic-e}}},
the second set follows directly from lines
\rnode{alg:atomic-b}{\ovalbox{\ref{alg:2:atomic-s}--\ref{alg:2:atomic-e}}},
while the \rnode{alg:sum-set-b}{\ovalbox{final}} specification
is the sum of all generated sets.
\nccurve[linestyle=dashed,angleA=20,angleB=320]{->}{alg:non-atomic-b}{alg:non-atomic-a}
\nccurve[linestyle=dashed,angleA=170,angleB=280]{->}{alg:atomic-b}{alg:atomic-a}
\nccurve[linestyle=dashed,angleA=90,angleB=335]{->}{alg:sum-set-b}{alg:sum-set-a}
Other examples are shown in Section~\ref{sec:reasoning-verification}.

\vspace{.5\baselineskip}
\noindent\textbf{Remarks and discussion.}
Let us supplement the approach and the Algorithm by some considerations.
The evidence from the approach is that atomic formulas $f1$ and $f4$ play an important role for every pattern.
They are always the first and the last, respectively, active activity/task for a pattern.
It means they constitute the entry and the exit, respectively, point for a pattern.
They are the first and the last, respectively, argument for a pattern.
In the work,
it is established that the maximum number of arguments for a pattern is $n=4$,
i.e.\ argument/formula $f_{4}$,
c.f.\ Fig.~\ref{fig:activity-workflow} and Fig.~\ref{fig:predefined-P}.
However, the limitation could be changed through introducing new perdefined patterns
which require, if necessary, greater number activities for a pattern.
Predefining new patterns,
one can introduce new types of iterations,
more complex forks or branchings,
and sequences with a greater number of activities
comparing the patterns in Fig.~\ref{fig:activity-workflow}.
These new patterns must have only one entry and only one exit argument.
Afterwards,
the new patterns must be defined in terms of temporal logic formulas,
c.f.\ Fig.~\ref{fig:predefined-P}.

The logical expression,
defined in Section~\ref{sec:generating-logical-specifications},
is similar to the well-known regular expression.
The internal structure of nested patterns (parentheses) is formed in such a way
that for any two patterns,
considered as arguments of the outer pattern, it is always satisfied that either
the first and the second patterns are completely disjointed,
or the first pattern is completely contained in the second one,
or the second patterns is completely contained in the first one.
It follows from the recursive nature of the logical expression definition
and correctly paired parenthesis.

The computational complexity of Algorithm~\ref{alg:generating-specification}
follows from the main loop which starts in line~\ref{alg:2:atomic-for}.
The number of patterns processed from left to right in the whole logical expression is expressed as a linear function.
If the considered pattern contains not only atomic arguments,
then it is necessary to calculate both joined entry and exit formulas.
On the other hand,
if non-atomic arguments do not take the most outer positions, i.e.\ $f1$ or $f4$, in the pattern,
that it is not necessary to go outside the pattern,
and it is enough to indicate the first or the last argument as joined entry or exit, respectively, formula.
Otherwise, the search for joined formulas follows from the current position to the left or right, respectively,
and in the worst-case to the beginning and to the end of the entire logical expression.
On the other hand, such worst-cases should be rare.
Summing up,
computational complexity is linearly dependent on the number of patterns in the logical expression,
and, in the worst-case, linearly from the calculation of joined formulas.
Thus, the worst-case complexity is bounded above by quadratic function.
For the average-case, the time complexity tends to the linear function.

A set of logical formulas is consistent if it does not contain contradiction,
i.e.\ it does not contain any two provable formulas such that the first formula is
a negation of the second formula.
As it has already been said,
logical patterns are predefined by a logician for further usage by an ordinary analyst.
This assumption is made and the logician is responsible for consistency of predefined specifications.
However, the open question is whether the Algorithm preserves consistency when generating logical specifications.
In the case of lines~\ref{alg:2:atomic-s}--\ref{alg:2:atomic-e},
due to the disjointedness of atomic formulas for patterns,
and the assumption of consistency of predefined patterns,
it is not possible to introduce contradictions when adding a new elementary set of formulas.
Let us note, that the following general formulas are valid and consistent:
$f1 \imp \som f4$, and $\alw\neg (f1 \con f4)$,
which means that if $f1$ (entry) is satisfied, then sometime in the future $f4$ (exit) is satisfied,
and that is always that $f1$ and $f4$ are not satisfied in the same time.
Due to the consistency of the above general formulas and formulas for every pattern,
the newly-generated logical specification is consistent,
and it follows from the fact that new temporal formulas for joined entry and exit formulas
refer to the consistent general/transition formulas of a pattern.

Completeness means that if a formula is true, it can be proven.
However, as it has already been said in the case of consistency,
the assumption is done that predefined specifications preserve completeness.
The question is whether the Algorithm preserves completeness when generating logical specification.
Let us note that all patterns in a logical expression are entirely nested,
i.e.\ it is not possible to obtain a partial nesting
that might provides an undesirable crossing of patterns.
Moreover, every two pattern contain disjoint sets of atomic formulas.
Completeness refers to the reachability all formulas and properties for every used logical pattern.
The entry and exit formulas are generalization for
a nested pattern and allow to bypass/skip its internal behaviour, if necessary,
in other words, they guarantee access to a pattern both
to/from the ``front'' and to/from the ``back'' of a pattern
with respect to both the preceding and the following pattern.
On the other hand, considering the disjunction of entry and exit formulas
for a pattern it allows to obtain/reach its entry end exit points, and consequently the remaining formulas.
It may cause some redundancy of generated formulas,
but on the other hand it covers all properties of combined patterns.
Thus, the Algorithm does not introduce itself incompleteness to the output logical specifications
with respect to predefined input specifications.

\section{Reasoning and verification}
\label{sec:reasoning-verification}

Let us summarize the entire method proposed in the work,
referring to the diagram at the end of Section~\ref{sec:methodology}.
The first phase,
let us call it the \emph{modeling phase},
enables development of requirements models
and includes the following steps:
\begin{itemize}
\item modeling of all use case diagrams $UCD_{1}, ..., UCD_{m}$,
      where $UC_{1}, ..., UC_{n}$ are all use cases contained in all use case diagrams;
\item modeling of scenarios for all use cases $UC_{1}, ..., UC_{n}$ and identification of
      atomic activities $AA=\{a_{1}, ..., a_{l}\}$;
\item modeling of activity diagrams $AD_{1}, ..., AD_{n}$ for all scenarios using
      predefined workflow patterns, c.f.\ Fig.~\ref{fig:activity-workflow},
      and using the identified atomic activities.
\end{itemize}
All the above steps require the assistance of an engineer and cannot be done automatically.
The next phase,
let us call it the \emph{analytical phase},
introduces a certain degree of automation and includes the following steps:
\begin{itemize}
\item translation of all activity diagrams $AD_{1}, ..., AD_{n}$ (and their workflows) to
      logical expressions $W_{L,1}, ..., W_{L,n}$;
\item generation of logical specifications $L_{1}, ..., L_{n}$
      for all logical expressions using the ${\cal A}$ algorithm,
      i.e.\ ${\cal A}(P,W_{L,i}) \longrightarrow L_{i}$ for every $i=1,...,n$;
\item summing of specifications, i.e.\ $L = L_{1} \cup ... \cup L_{n}$;
\item (manual) definition of the desired property $Q$;
\item start of the process of automatic reasoning using the semantic tableaux method
      for formula $f_{1} \con ... \con f_{k} \imp Q$, where $f_{1}, ..., f_{k}$
      are formulas which belong to the logical specification $L$.
\end{itemize}
The above steps illustrate the entire operation of the system shown in
Fig.~\ref{fig:deduction-system}.
The loop between the last two steps,
c.f.\ Fig.~\ref{fig:formalization},
refers to a process of
both introducing and verifying more and more new properties (formula $Q$) of
the examined model.

Let us consider the activity diagrams $AD_{2}$ and $AD_{3}$ shown in Fig.~\ref{fig:activity-diagram}
for use cases $UC_{2}$ ``InsuranceDamageLiquidation'' and $UC_{3}$ ``VehicleCarriage''.
Activity diagrams constitute the input for the deduction system shown in Fig.~\ref{fig:deduction-system}.
The logical expression $W_{L,2}$ for $AD_{2}$ is
\begin{eqnarray*}
\begin{array}{l}
    Seq(SystemLogIn,Branch(DamageVindication,Concur(\\
    Seq(SupplyDocumentaryEvidence,InitRepair),\\
    Seq(MechanicalRepairs,BodyRepairs),RentVehicle,TestDrive),\\
    WarningDoubleVindication,SystemLogOut))
\end{array}
\end{eqnarray*}
The logical expression $W_{L,3}$ for $AD_{3}$ is
\begin{eqnarray*}
\begin{array}{l}
    Seq(Seq(CallAssistance,TransportVehicle),InitRepair)
\end{array}
\end{eqnarray*}
Substituting letters of the Latin alphabet in places of propositions:
$a$ -- Application,
$b$ -- DamageVindication,
$c$ -- SupplyDocumentaryEvidence,
$d$ -- InitRepair,
$e$ -- MechanicalRepairs,
$f$ -- BodyRepairs,
$g$ -- RentVehicle,
$h$ -- TestDrive,
$i$ -- WarningDoubleVindication,
$j$ -- FinalReport,
$k$ -- CallAssistance, and
$l$ -- TransportVehicle,
then the expression $W_{L,2}$ is
\begin{eqnarray}
   Seq(a,Branch(b,Concur(Seq(c,d),Seq(e,f),g,h),i,j))\label{formula:logical-expression-1}
\end{eqnarray}
and the expression $W_{L,3}$ is
\begin{eqnarray}
   Seq(Seq(k,l),d)\label{formula:logical-expression-2}
\end{eqnarray}
Replacing propositions (atomic activities) by Latin letters is a technical matter
and is suitable only for the work because of its limited size.
In the real world,
c.f.\ the deduction system from Fig.~\ref{fig:deduction-system},
original names of the activities would be used.

A logical specification $L_{3}$ for the logical expression $W_{L,3}$ is built in some steps
that result from Algorithm~\ref{alg:generating-specification}.
At the beginning, the specification of a model is $L_{3}=\emptyset$.
Patterns are processed from left to right.
Hence, the first processed pattern is $Seq$ and the next considered are:
$Branch$, $Concur$, $Seq$, and $Seq$.
The resulting logical specifications contain the formulas
\begin{eqnarray}
L_{2}=\{
a \imp \som (b \dis j),
\neg a \imp \neg\som (b \dis j),
\alw\neg(a \con (b \dis j)),\nonumber\\
b \imp (\som (c \dis h) \con \neg\som i) \dis (\neg\som (c \dis h) \con \som i),\nonumber\\
\neg b \imp \neg((\som (c \dis h) \con \neg\som i) \dis (\neg\som (c \dis h) \con \som i)),\nonumber\\
(c \dis h) \dis i \imp \som j,
\neg((c \dis h) \dis i) \imp \neg\som j,
\alw\neg(b \con j),
\alw\neg((c \dis h) \con i),\nonumber\\
\alw\neg(b \con ((c \dis h) \dis i)),
\alw\neg(((c \dis h) \dis i) \con j),
(c \dis d) \imp \som (e \dis f) \con \som g,\nonumber\\
\neg (c \dis d) \imp \neg(\som (e \dis f) \con \som g),
(e \dis f) \con g \imp \som h,
\neg((e \dis f) \con g) \imp \neg\som h,\nonumber\\
\alw\neg((c \dis d) \con ((e \dis f) \dis g)),
\alw\neg(((e \dis f) \dis g) \con h),
\alw\neg((c \dis d) \con h),\nonumber\\
c \imp \som d,
\neg c \imp \neg\som d,
\alw\neg(c \con d),
e \imp \som f,
\neg e \imp \neg\som f,
\alw\neg(e \con f)
\}  \label{formula-example-specification-2}
\end{eqnarray}
and
\begin{eqnarray}
L_{3}=\{
(k \dis l) \imp \som d,
\neg(k \dis l) \imp \neg\som d,
\alw\neg((k \dis l) \con d),\nonumber\\
k \imp \som l,
\neg k \imp \neg\som l,
\alw\neg(k \con l)
\}  \label{formula-example-specification-3}
\end{eqnarray}
Formulas~\ref{formula-example-specification-2} and~\ref{formula-example-specification-3}
represent the output,
i.e.\ $L_{2} \cup L_{3} = L$,
of the \psframebox[boxsep=true,shadow=false]{\scriptsize G} component in Fig.~\ref{fig:deduction-system}.

Formal \emph{verification} is the act of proving the correctness of a system (liveness, safety).
\emph{Liveness} means that the computational process achieves its goals,
i.e.\ something good eventually happens.
\emph{Safety} means that the computational process avoids undesirable situations,
i.e.\ something bad never happens.
The liveness property for the model can be
\begin{eqnarray}
b \imp\som g \label{examined-property-liveness-1}
\end{eqnarray}
which informally/verbally means that
\textbf{if the damage vindication is satisfied then sometime in the future the replacement car is reached},
formally\ $DamageVindication \imp\som RentVehicle$.
Another example, including extended use case, of the liveness property is
\begin{eqnarray}
k \imp\som h \label{examined-property-liveness-2}
\end{eqnarray}
which means that
\textbf{if the call assistance is satisfied then sometime in the future the test drive is reached},
formally\ $CallAssistance \imp\som TestDrive$.
The safety property for the examined model can be
\begin{eqnarray}
\alw\neg (i \con g) \label{examined-property-safety-1}
\end{eqnarray}
which means that
\textbf{it never occurs that the rental of a vehicle and the double vindication are satisfied in the same time},
or more formally\ $\alw\neg (WarningDoubleVindication \con RentVehicle)$.
Another example of the safety property is
\begin{eqnarray}
\alw\neg (i \con l) \label{examined-property-safety-2}
\end{eqnarray}
which means that
\textbf{it never occurs that the transport of a vehicle and the double vindication are satisfied in the same time},
or more formally\ $\alw\neg (WarningDoubleVindication \con TransportVehicle)$.
When considering the property expressed by Formula~\ref{examined-property-liveness-1},
then the whole formula to be analyzed using semantic tableaux,
providing a combined input for
the \psframebox[boxsep=true,shadow=false]{\scriptsize T} component
in Fig.~\ref{fig:deduction-system}, is
\begin{eqnarray}
(a \imp \som (b \dis j)) \con ..... \con \alw\neg(k \con l)
\imp (b \imp\som g)\quad \label{formula-example-property-liveness-1}
\end{eqnarray}
where the antecedent of the (main) implication is the conjunction of formulas from
the generated sets $L_{2}$ and $L_{3}$ (Formulas~\ref{formula-example-specification-2} and~\ref{formula-example-specification-3}),
i.e.\ conjunction $C(L)=C(L_{2} \cup L_{3})$.
In a similar way, the whole formula for Formula~\ref{examined-property-liveness-2}
is expressed by
\begin{eqnarray}
C(L_{2} \cup L_{3})
\imp (k \imp\som h)\quad \label{formula-example-property-liveness-2}
\end{eqnarray}
When considering the property expressed by
Formulas~\ref{examined-property-safety-1} and~\ref{examined-property-safety-2},
then the whole formulas are constructed in a similar way as
\begin{eqnarray}
C(L_{2} \cup L_{3})
\imp \alw\neg (i \con g) \label{formula-example-property-safety-1}
\end{eqnarray}
and
\begin{eqnarray}
C(L_{2} \cup L_{3})
\imp \alw\neg (i \con l) \label{formula-example-property-safety-2}
\end{eqnarray}
In all cases,
i.e.\ Formulas~\ref{formula-example-property-liveness-1},
\ref{formula-example-property-liveness-2},
\ref{formula-example-property-safety-1},
and~\ref{formula-example-property-safety-2},
after the negation of the input formula within the prover, the inference trees are built.
Presentation of a full inference tree for both cases exceeds the size of the work.
(The simple inference tree from Fig.~\ref{fig:inference-tree} gives an idea how it works.)
All branches of the semantic trees are closed,
i.e.\ Formulas from~\ref{examined-property-liveness-1} to~\ref{examined-property-safety-2}
are satisfied in the considered requirements model.
In the case of falsification of the semantic tree
the open branches are obtained and provide information about
the source of an error what is another advantage of the method.

Although the logical specification was generated for only activity diagrams $AD_{2}$ and $AD_{3}$,
that is $L=L_{2} \cup L_{3}$, c.f.\ Formula~\ref{formula-example-specification-2} and~\ref{formula-example-specification-3},
the method is easy to scale up, i.e.\ extending and summing up logical specifications
for other activity diagrams and their scenarios.
Then, it will be possible to examine logical relationships (liveness, safety) for
different activities coming from different activity diagrams
for different use cases and scenarios.

\section{Conclusions}
\label{sec:conclusion}

The work proposes a two-phase strategy for formal analysis of requirements models.
The first one is carried out by an engineer using a defined methodology
and the second one can be (in most) automatic and enables formal verification of
the desired properties (liveness, safety) of behaviour.
The method for an automatic generation of logical specifications is proposed.
This specification is a set of temporal logic formulas and obtaining it
is crucial in the case of a practical use of the deduction-based formal verification.

The proposed method enables the construction in a formal way requirements models
and then extracting logical specifications.
The proposed method of generating enables scaling up,
when building more and more nesting patterns.
Logical patterns are once defined (logician) and could be commonly used by users (engineers).
Introducing logical patterns as logical primitives allows for
breaking of some barriers and obstacles in receiving logical specifications as
a set of a large number of temporal logic formulas in an automated way.
Application of formal verification, which is based on deductive inference,
helps to significantly increase the maturity of requirements models
considering infinite computations and using a human-intuitive approach.

Future works may include the implementation of the logical specification generation module.
Another important issue could be a detailed analysis of the available provers~\cite{Schmidt-2013-provers}
which could be useful and applied for the approach.
It should result in a CASE software,
e.g.\ Integrated Development Environments (IDEs),
which could be a first step involved in creating industrial-proof tools,
i.e.\ implementing another part of formal methods, hope promising, in industrial practice.
The literature review argues that there is a lack of such comprehensive tools.

\bibliographystyle{splncs03}
\bibliography{../bib/rk-bib-rk,../bib/rk-bib-main,../bib/rk-bib-requirements,../bib/rk-bib-inne}

\end{document}